\documentclass[twocolumn,showpacs,preprintnumbers,amsmath,amssymb]{revtex4}

\newcommand{\maru}[1]{\left( #1 \right)}

\newcommand{\nami}[1]{\left\{ #1 \right\}}
\def\f{\frac}

\def\be{\begin{equation}}
\def\ee{\end{equation}}

\def\be{\begin{equation}}
\def\ee{\end{equation}}

\def\fr{\frac}

\def\be{\begin{equation}}
\def\ee{\end{equation}}

\def\fr{\frac}

\def\red{\textcolor{black}}
\def\aka{\textcolor{black}}
\def\blue{\textcolor{black}}
\def\ao{\textcolor{black}}
\def\sea{\textcolor{black}}
\def\umi{\textcolor{black}}
\def\cyan{\textcolor{black}}
\def\cyann{\textcolor{black}}
\def\magenta{\textcolor{black}}
\def\green{\textcolor{black}}
\usepackage[dvipdfmx]{graphicx}
\usepackage{comment}
\usepackage{epsf}
\usepackage{amsmath}
\usepackage{graphics}
\usepackage{amsfonts}
\usepackage{amssymb}
\usepackage{latexsym}
\usepackage{color}
\input{colordvi.tex}
\usepackage{feynmp}

\begin{document}
\preprint{RESCEU-20/15}

\title{
\red{
Primordial black holes as a novel probe of \\primordial gravitational waves
}
}
\author{Tomohiro Nakama$^{1,2}$}
\author{Teruaki Suyama$^{2}$}
\affiliation{
$^1$
  Department of Physics, Graduate School of Science,\\ The University of Tokyo, Tokyo 113-0033, Japan
}
\affiliation{
$^2$
  Research Center for the Early Universe (RESCEU), Graduate School
  of Science,\\ The University of Tokyo, Tokyo 113-0033, Japan
}

\date{\today}

\begin{abstract}
We propose a novel method to probe primordial gravitational waves 
by means of primordial black holes (PBHs). 
When the amplitude of \red{primordial} tensor perturbations 
on comoving scales much smaller than those relevant to Cosmic Microwave Background is very large, 
it induces scalar perturbations due to second-order effects \red{substantially}. 
If the amplitude of resultant scalar perturbations becomes too large, 
then \red{PBHs are overproduced} to a level that is inconsistent with 
a variety of existing observations constraining the\aka{ir} abundance. 
This leads to upper bounds on the amplitude of initial tensor perturbations \cyan{on super-horizon scales}. 
These upper bounds from PBHs are compared with other existing bounds. 
\end{abstract}


\maketitle
\textit{Introduction.-}
\red{
Stochastic gravitational wave background (SGWB) 
on a wide range of scales is thought to have been generated in the early universe. 
}
SGWB on \umi{the} largest observable scales have been investigated by Planck \cite{Ade:2013zuv} and BICEP2 \cite{Ade:2014xna}. 
SGWB on smaller scales can be constrained 
by inferring the value of $N_{\rm{eff}}$, 
the effective number of extra degrees of freedom of relativistic species, 
at Big Bang Nucleosynthesis (BBN) through the current abundance of light elements \cite{Allen:1996vm}, 
or at photon decoupling through the anisotropy of Cosmic Microwave Background (CMB) \cite{Smith:2006nka,Kikuta:2014eja}. 
More recently SGWB on small scales has \aka{been} \cyan{constrained} by CMB spectral distortions as well \cite{Ota:2014hha,Chluba:2014qia}. 
SGWB can also be \aka{directly} constrained by gravitational wave detectors (see e.g. \cite{Aasi:2014zwg}). 
BBN and CMB have played a major role in constraining SGWB since they are applicable on a wide range of scales, 
noting ground-based laser interferometers tend to target gravitational waves (GWs) on a relatively limited frequency range with high sensitivity. 
However, it would be worthwhile to note that upper bounds obtained through $N_{\rm{eff}}$ \umi{need} an assumption about 
the number of relativistic species in the early universe, as is discussed later. 
In addition, in obtaining BBN or CMB bounds we implicitly assume that 
any physical mechanisms, both known and unknown, \textit{increase} $N_{\rm{eff}}$, 
making $N_{\rm{eff}}$ larger than the standard value $N_{\rm{eff}}=3.046$ \cite{Mangano:2005cc}. 
However, in principle it is possible to \textit{decrease} $N_{\rm{eff}}$ (see e.g. \cite{Ichiki:2002eh,Apostolopoulos:2005at,Maartens:2010ar}).
\red{Therefore, }
it would be desirable to have another independent cosmological method to probe 
SGWB on a wide range of scales, which does not \umi{much} depend on the assumptions mentioned above. 

The very physical mechanism we employ to probe SGWB is the formation of 
primordial black holes (PBHs), black holes formed in the early universe well before the \red{cosmic} structure formation. 
\blue{There are a number of possible} mechanisms to create PBHs (see e.g. \cite{Carr:2009jm} and references therein), 
but one of the simple and natural mechanisms is the direct collapse of radiation overdensity 
during \red{the radiation-dominated era}, which happens when the density perturbation becomes order unity 
at the moment of the horizon crossing of the perturbation \cite{Zel'dovich-1974,Carr:1974nx,Carr:1975qj}. 
(See also \cite{Harada:2013epa,Harada:2015yda} for updated discussions of the formation condition and see also 
\cite{1978SvA....22..129N,Shibata:1999zs,Niemeyer:1999ak,Polnarev:2006aa,Polnarev:2012bi,Nakama:2013ica,Nakama:2014fra} for numerical simulations of the formation of PBHs.) 
There is no conclusive evidence for the existence of PBHs and so upper bounds on the 
abundance of PBHs on various mass scales have been obtained by various kinds of observations (see e.g. \cite{Carr:2009jm} and references therein). 
These upper bounds can be translated into upper bounds on the power spectrum of the curvature perturbation on small scales \cite{Bugaev:2008gw,Josan:2009qn}, 
namely, they can be used to exclude models of the early universe which predict too many PBHs. 
(Other methods to constrain primordial perturbations on small scales include CMB spectral distortions \cite{Chluba:2012we}, 
acoustic reheating \cite{Nakama:2014vla,Jeong:2014gna}, ultracompact minihalos \cite{Bringmann:2011ut,Kohri:2014lza} \aka{etc}.) 

\red{In this paper we use }PBHs to constrain \red{primordial tensor fluctuations}. 
\red{Large} amplitude tensor perturbations induce scalar perturbations (\textit{induced scalar perturbations}) due to 
their second-order effects, though tensor and scalar perturbations are decoupled at the linear level. 
If the typical amplitude of primordial GWs is too large, then the typical amplitude of 
resultant induced scalar perturbations becomes also too large and there appear too many regions where 
the amplitude of density perturbation becomes order unity at the horizon crossing to form PBHs. 
\aka{Thus} we can obtain upper bounds on the amplitude of \aka{primordial} tensor modes requiring PBHs are not overproduced \aka{as they enter the Hubble radius}
\footnote{
The direct gravitational collapse of non-linear localized gravitational waves has been discussed in the literature 
\cite{Eppley:1977dk,Miyama:1981mh,Shibata:1993fx,Shibata:1995we,Shibata:1997ix,Alcubierre:2000xu} 
and so one may place upper bounds on tensor modes \blue{also} using this direct collapse mechanism. 
However, the initial condition and dynamics of non-linear gravitational waves originated from SGWB during the \red{radiation-dominated era} 
have not been fully understood and so in this paper we consider the \red{second-order GWs}, 
noting the dynamics of non-linear radiation density perturbations is better understood. 
}. 

\red{
\cyan{Due to our current ignorance of the correct model of inflation and the subsequent thermal history of the universe, 
new upper limits on tensor perturbations on small scales in themselves are worthwhile. }
In addition, it makes our new upper limits still more valuable that 
there are models of the early universe which 
can predict large tensor perturbations on small scales \cite{Khoury:2001wf,Baldi:2005gk,Brandenberger:2006xi,Mielczarek:2010bh,Kobayashi:2010cm,Kobayashi:2011nu,Gong:2014qga,Mukohyama:2014gba,Ashoorioon:2014nta,Biswas:2014kva,Qiu:2015nha}. 
If a model predicts large tensor perturbations on small scales, and even larger scalar perturbations at the same time, 
then such a model would be more severely constrained by PBHs generated from first-order scalar perturbations. 
Here we consider PBH formation only from second-order tensor perturbations, but if scalar perturbations are also large, PBHs are formed more, and so 
our bounds on tensor perturbations are conservative or model-independent, in the sense that those bounds do not depend on the amplitude of \blue{first-order} scalar perturbations on small scales. 
Note also that some of the above previous works can (or possibly can) predict not only large tensor perturbations on small scales, but also large \textit{tensor-to-scalar ratio} on small scales, and our method is particularly 
useful to constrain these types of models. 
}

To calculate induced scalar perturbations one may solve the Einstein equations for 
scalar perturbations including the source terms which are second-order in tensor perturbations, 
assuming some form of the initial power spectrum of tensor perturbations. 
In general, the statistics of induced scalar perturbation\red{s are} highly non-Gaussian, 
since \red{they are} generated by second-order tensor perturbations. 
If we assume some statistical properties of initial tensor perturbations, e.g. Gaussianity, 
in principle one can determine the probability density function \red{(PDF)} of induced scalar perturbation, 
which is necessary to calculate the abundance of PBHs given some initial tensor power spectrum 
to obtain PBH constraints on \red{primordial} GWs. 
All the details of those analyses will be \red{presented} elsewhere \cite{prep}, 
and here we 
present simple analytic estimations,
which correctly reproduce the main results of \cite{prep}, 
as is discussed shortly.
\begin{figure}[t]
\begin{center}
\includegraphics[width=8.5cm,keepaspectratio,clip]{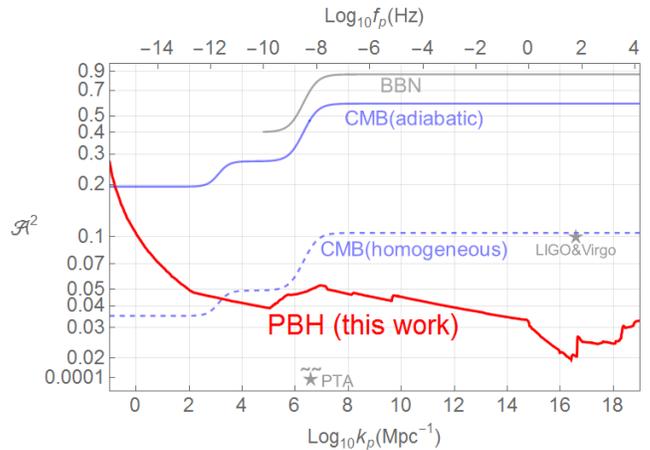}
\end{center}
\caption{\red{Comparison of the constraints on the amplitude of the delta-function power spectrum eq.(\ref{delta}), obtained from PBH, BBN, CMB, \blue{PTAs,} LIGO and Virgo. 
For more details see \cite{prep}. \cyann{\umi{The} PBH constraints in this figure are obtained assuming 
Gaussianity of primordial tensor perturbations and calculating the PDF of induced radiation perturbations numerically in \cite{prep}. 
The thick and dashed blue lines correspond to CMB constraints for adiabatic and homogeneous initial conditions of GWs, respectively. 
Note that the constraint from PTA is more than two orders of magnitude tighter than those from PBHs at $k_p\sim 4\times 10^6\mathrm{Mpc}^{-1}$.
The frequency corresponding to each $k_p$, $f_p=ck_p/2\pi$, is also shown. 
See the text for more details. }
} 
}
\label{constraints}
\end{figure}

\textit{PBH upper bound\red{s} on tensor perturbation.-}
For simplicity we assume the typical amplitude of tensor fluctuations 
of only a limited range of \umi{wavelengths} \red{around $\lambda_p$} (say, in the range $(e^{-1/2}\lambda_p,e^{1/2}\lambda_p)$) is large\red{ ,
namely we consider a peak in the power spectrum around $\lambda_p$}. 
Let us denote by $\sigma_h$ the \red{root-mean-square amplitude} (RMS) or \red{the} typical amplitude of tensor fluctuations 
when the relevant modes under consideration are on super-horizon scales. 
Note that $\sigma_h^2$ roughly corresponds to the dimensionless power spectrum 
${\cal P}_h$ \blue{of primordial GWs}. 
Suppose \red{these large amplitude tensor perturbations whose wavelength is around $\lambda_p$ reenter} the horizon during \red{the radiation-dominated era}. 
It generates radiation density perturbations $\delta_r$ \red{whose wavelength is around $\lambda_p$} and whose RMS $\sigma_{\delta_r}$ 
at around the horizon crossing is roughly given by 
$\sigma_{\delta_r}\sim \sigma_h^2$, simply because $\delta_r$ is generated by \red{second-order GWs}. 
\red{As confirmed in \cite{prep}, this relation indeed holds well at least for the delta-function power spectrum. }
If $\delta_r>\delta_{r,\mathrm{th}}\sim \red{0.4}$ \red{\cite{Harada:2013epa}}, 
the induced radiation density perturbations collapse to 
form PBHs. 
The probability of \umi{the} formation of PBHs has to be extremely small to be consistent with 
various observations \cite{Carr:2009jm}. 
Therefore, in order not to overproduce PBHs, one may require
${\cal P}_h\sim\sigma_h^2\sim\sigma_{\delta_r}\lesssim 0.1\delta_{r,\mathrm{th}}\sim 0.0\red{4}$. 
\blue{More comments will be made later regarding this inequality. }
This upper bound is applicable on all the scales which reenter the horizon during \red{the radiation-dominated era}. 
Furthermore, this bound is \aka{probably} conservative on scales which reenter the horizon during the (early as well as late) matter\red{-dominated era}, 
since formation of black holes by direct collapse of primordial perturbations is easier when 
the universe is (literally or effectively) dominated by pressureless dust 
\aka{
\footnote{
\aka{
Note that PBH formation 
due to second-order GWs during a dust-dominated universe
can not be completely regarded as 
collapse of dust in a dust-dominated universe, 
since locally the energy density of GWs is sizable, 
and so the effective equation of state there would be positive.
}
}
}. 
To summarize, we have 
\begin{equation}
{\cal P}_h\lesssim 0.0\red{4} \quad(\mathrm{PBH}),\label{PBH}
\end{equation}
and this is applicable from the comoving wavelength of $\sim 1\mathrm{Gpc}$ 
all the way down to $\sim 1\mathrm{Gpc}\times e^{-60}\sim 0.3$m, assuming the total number of e-foldings during inflation is sixty 
\footnote{
\red{The idea of using PBHs to constrain tensor fluctuations itself is so simple and indeed it does not seem to be new. 
For example, we found(, when finalizing our manuscript,) in \cite{Wang:2014kqa} 
PBH constraints are briefly mentioned. There they require ${\cal P}_h\lesssim 1$, which is 
probably too conservative to ensure the sufficient rareness of formation of PBHs to be consistent with observations (see eq.(\ref{PBH}) and the \blue{preceding arguments}). 
We discuss PBH constraints on tensor fluctuations in detail, quantitatively in this paper and an accompanying paper \cite{prep} for the first time. 
}
}. 
See \textit{Discussion} for some comments regarding the constraints on smallest scales. 

\blue{Our analytic estimations} can reproduce the results of more rigorous calculations in \cite{prep}, 
where the delta-function initial tensor power spectrum is assumed: 
\begin{equation}
{\cal P}_h(k)={\cal A}^2k_p\delta(k-k_p).\label{delta}
\end{equation} 
The PBH abundance for this power spectrum is calculated and constraints on ${\cal A}^2$ as a function of $k_p$ are obtained, 
which is shown in Fig.\ref{constraints}, along with other constraints. 
One can see that \blue{our analytic estimations} (eqs. (\ref{PBH}), (\ref{BBN}) and (\ref{CMB})) indeed explain Fig.1 well. 

\textit{Comparison with other methods.-}
First note that the total energy density of radiation without the presence of gravitational waves can be written as follows;
\be
\rho_{\rm{rad}}(T)=\red{(\pi^2/30) g_*}T^4. 
\ee
\red{Here $g_*$} is 
the effective number of degrees of freedom of relativistic species \red{and is} given by\red{, before the electron-positron annihilation,} \cite{Allen:1996vm,Maggiore:1999vm}
\be
\red{g_*}=2+\fr{7}{8}(4+2N_\nu),
\ee
where $N_\nu$ is the effective number of degrees of freedom of neutrinos $N_\nu=3.046$. 
\red{It is convenient to denote the presence of GWs (or possibly dark radiation) by $\Delta N_{\rm{eff}}$, as a correction to $N_{\nu}$ above \cite{Allen:1996vm}}. 
In the following we use $\Delta N_{\rm{GW}}$ as a contribution of GWs and derive the expression of it in terms of ${\cal P}_h$, 
enabling observational upper limits on $\Delta N_{\rm{eff}}$ to be translated into upper limits on ${\cal P}_h$.
When GWs are present, the critical density $\rho_{\rm{crit}}$ during the \red{radiation-dominated era} is 
\be
\rho_{\rm{crit}}\simeq\rho_{\rm{rad}}+\rho_{\rm{GW}},
\ee
where $\rho_{\rm{rad}}$ and $\rho_{\rm{GW}}$ are the energy density of radiation and GWs respectively.
After the horizon crossing of GWs, their energy density starts to scale as $\propto a^{-4}$,  
while, denoting the effective degrees of freedom of relativistic species in terms of entropy at temperature $T$ by $g_S(T)$, 
the photon temperature evolves following $g_S(T)T^3a^3=$const. (i.e. constant entropy) and so $\rho_{\rm{rad}}\red{\propto g_*/(a^4g_S^{4/3})\sim 1/(a^4g_S^{1/3})}$ 
(see e.g. \cite{Maggiore:1999vm}). 
Let us define $\Omega_{\rm{GW}}\equiv\rho_{\rm{GW}}/\rho_{\rm{crit}}\simeq \rho_{\rm{GW}}/\rho_{\rm{rad}}$.
Then, 
\be
\Omega_{\rm{GW}}(T)=\left(\frac{g_S(T)}{g_S(T_{\rm{in}})}\right)^{\red{1}/3}\Omega_{\rm{GW}}(T_{\rm{in}}),\label{omegagw}
\ee
where $T_{\rm{in}}$ is the temperature of radiation at the moment of the horizon crossing of GWs \red{whose wavelength is $\sim \lambda_p$}, and 
$T<T_{\rm{in}}$.
\cyann{At the epoch of BBN, t}he contribution of GWs is characterized by $\Delta N_{\rm{GW}}$ as follows;
\begin{align}
&\rho_{\rm{rad}}(T)+\rho_{\rm{GW}}(T)\nonumber\\
&=\cyann{\frac{\pi^2}{30}}
\left[
2+\frac{7}{8}
\left\{
4+2(N_{\rm{\red{\nu}}}+\Delta N_{\rm{GW}})
\right\}
\right]T^4,\label{total}
\end{align}
which leads to
\begin{align}
&\rho_{\rm{GW}}(T)\nonumber\\
&=\rho_{\rm{rad}}(T)\times\frac{7}{8}\times 2\times \Delta N_{\rm{GW}}(T)/
\left\{
2+\frac{7}{8}(4+2N_{\rm{\red{\nu}}})
\right\}&\nonumber\\
&\simeq\frac{7}{43}\rho_{\rm{rad}}\red{(T)}\Delta N_{\rm{GW}}(T).&
\end{align}
Noting 
\footnote{
 \red{
      For the delta-function power spectrum eq.(\ref{delta}) one can actually show that $\Omega_{\rm{gw}}(T_{\rm{in}})=2{\cal A}^2/3$ \cite{prep}, 
      \cyann{and this relation is used to obtain Fig.1. }
     }
}
\be
\Omega_{\rm{GW}}(T_{\rm{in}})\sim {\cal P}_h,
\ee
$\Delta N_{\rm{GW}}(T)$ can be written as
\be
\Delta N_{\rm{GW}}(T)=\fr{43}{7}\Omega_{\rm{GW}}(T)\sim\frac{43}{7}{\cal P}_h
\left(\frac{g_S(T)}{g_S(T_{\rm{in}})}\right)^{\red{1}/3}.\label{DeltaNGW}
\ee
\red{In the literature a}n upper bound on $\Delta N_{\rm{eff}}<\Delta N_{\rm{upper}}$ is translated into an upper bound on 
$\Delta N_{\rm{GW}}$, $\Delta N_{\rm{GW}}<\Delta N_{\rm{upper}}$. 
\red{As is already mentioned}, 
here it is assumed that any physical mechanisms, both known and unknown, contribute positively to $N_{\rm{eff}}$. 
However, it would be worthwhile to note that at least there are examples where $N_{\rm{eff}}$ decreases \cite{Ichiki:2002eh,Apostolopoulos:2005at,Maartens:2010ar}. 
With this in mind, the requirement above, $\Delta N_{\rm{GW}}<\Delta N_{\rm{upper}}$, is translated into an upper bound on ${\cal P}_h$ from (\ref{DeltaNGW}) as follows:
\be
{\cal P}_h<\fr{7}{43}\Delta N_{\rm{upper}}\left(\frac{g_S(T_{\rm{in}})}{g_S(T)}\right)^{\red{1}/3}.
\ee
On the other hand, at the epoch of photon decoupling,
\begin{align}
&\rho_{\rm{rad}}(T)+\rho_{\rm{GW}}(T)\nonumber\\
&=\frac{\pi^2}{30}\nami{2+2\times \frac{7}{8}\maru{\frac{4}{11}}^{4/3}(N_\nu+\Delta N_{\rm{GW}})},
\end{align}
which yields
\begin{equation}
\Omega_{\rm{GW}}(T)=\frac{2\times\f{7}{8}\maru{\frac{4}{11}}^{4/3}}{2+2\times\f{7}{8}\maru{\frac{4}{11}}^{4/3}N_\nu}\Delta N_{\rm{GW}}
\simeq 0.13\Delta N_{\rm{GW}}.
\end{equation}
So in this case we find
\be
{\cal P}_h<0.13\Delta N_{\rm{upper}}\left(\frac{g_S(T_{\rm{in}})}{g_S(T)}\right)^{1/3}.
\ee
Note that this constraint depends on $g_S$, 
which one may \red{regard as} a drawback of these methods 
since $g_S$ is uncertain especially at high temperatures. 
It also depends on other potential entropy productions \cite{Kuroyanagi:2014nba}. 
On the other hand, the PBH constraint does not \umi{much} depend on $g_S$ nor other entropy productions. 
Assuming the standard model of particle physics, we use the following for simplicity:
\be
g_S(T)\sim
\left\{
\begin{array}{ll}
100 , &\quad (150\rm{\red{M}eV}\lesssim \textit{T})  \\
11 , &\quad (1\rm{MeV}\lesssim \textit{T}\lesssim 150\rm{\red{M}eV})  \\
4 , &\quad (T\lesssim 1\rm{MeV})
\end{array}
\right.
\ee
Note that $T\sim 150\rm{\red{M}eV}$ corresponds to the QCD phase transition, and 
$T\sim 1$MeV to the electron-positron annihilation. 
In order not to spoil BBN, we set, following \cite{Kuroyanagi:2014nba}, 
$\Delta N_{\rm{upper}}=1.65$
as a 95\% C.L. upper limit, 
which \red{can constrain GWs only of} the scales smaller than the comoving horizon at the time of BBN, namely, $70\rm{kpc}^{-1}\lesssim \textit{k}$. 
Noting the comoving scales which enter the horizon at $T\sim 150$\red{M}eV and $T\sim 1$MeV correspond to 
$k\sim 4\rm{pc}^{-1}$ and $k\sim 1\rm{kpc}^{-1}$ respectively, 
we obtain the following constraints 
\be
{\cal P}_h\lesssim
\left\{
\begin{array}{ll}
\cyann{0.56} , &\quad (4\rm{pc}^{-1}\lesssim \textit{k})  \\
\cyann{0.27} , &\quad (70\rm{kpc}^{-1}\lesssim \textit{k}\lesssim 4\rm{pc}^{-1})  
\end{array}
\right.\label{BBN}
\ee

As for CMB constraints, 
in \cite{Smith:2006nka} the use of homogeneous initial conditions of GWs is advocated for 
SGWB generated, for instance, by quantum fluctuations during inflation. 
In this case 95 \% upper limits are $\Delta N_{\rm{upper}}=0.18$ \cite{Sendra:2012wh} \sea{\footnote
{
\sea{
One would get somewhat tighter constraints than those in \cite{Sendra:2012wh} for 
homogeneous initial conditions of GWs energy density, 
by repeating the analysis of \cite{Sendra:2012wh} using more recent data.
}
}
}, 
which correspond to 
\be
{\cal P}_h\lesssim
\left\{
\begin{array}{ll}
\cyann{0.068} , &\quad (4\rm{pc}^{-1}\lesssim \textit{k})  \\
\cyann{0.033} , &\quad (1\rm{kpc}^{-1}\lesssim \textit{k}\lesssim 4\rm{pc}^{-1})  \\
\cyann{0.023} , &\quad (10\rm{Gpc}^{-1}\lesssim \textit{k}\lesssim 1\rm{kpc}^{-1})  
\end{array}
\right.\label{CMBhomo}
\ee
\sea{For adiabatic initial conditions of GWs we refer to
\begin{equation}
N_{\rm{eff}}=3.52_{-0.45}^{+0.48}\quad(95\%;\;\;Planck+\rm{WP}+\rm{highL}+H_0+\rm{BAO})
\end{equation}
of \cite{Ade:2013zuv} to set $\Delta N_{\rm{upper}}=1.00$ \cite{Kikuta:2014eja}:
} 
\be
{\cal P}_h\lesssim
\left\{
\begin{array}{ll}
\cyann{0.38} , &\quad (4\rm{pc}^{-1}\lesssim \textit{k})  \\
\cyann{0.18} , &\quad (1\rm{kpc}^{-1}\lesssim \textit{k}\lesssim 4\rm{pc}^{-1})  \\
\cyann{0.13} , &\quad (10\rm{Gpc}^{-1}\lesssim \textit{k}\lesssim 1\rm{kpc}^{-1})  
\end{array}
\right.\label{CMB}
\ee
One may not regard these constraints as meaningful, because upper limits correspond to 
the amplitude of GWs which is (almost) non-linear. 

The current energy density of SGWB, $\Omega_{\rm{GW},0}$, is \red{also constrained by LIGO and Virgo, most severely 
\magenta{in the band $41.5-169.25\rm{Hz}$} as $\Omega_{\rm{GW},0}\lesssim 5.6\times 10^{-6}\times\log(169.25/41.5)\simeq 8\times 10^{-6}$ \cite{Aasi:2014zwg}}. 
Noting $\Omega_{\rm{GW},0}\sim z_{\rm{eq}}^{-1}(4/100)^{\red{1}/3}{\cal P}_h\sim 10^{-4}{\cal P}_h$ 
($z_{\rm{eq}}\sim 3000$ is the redshift at the matter-radiation equality, and \red{the factor $z_{\rm{eq}}^{-1}$} reflects $\Omega_{\rm{GW}}\propto (1+z)/(1+z_{\rm{eq}})$ during \red{a matter-dominated era}), 
we have ${\cal P}_h\lesssim \magenta{0.08}$ 
\cyan{
 \footnote{
  \cyan{
  Note that they also obtained weaker upper bounds on a few other frequency ranges other than the one around $\sim 100$Hz.
  Strictly speaking in \cite{Aasi:2014zwg} some power low frequency dependence is assumed in each band, 
  and so their results may not be directly translated into constraints on \blue{a narrow peak in the} power spectrum we consider. 
  \ao{Indeed in \cite{Nishizawa:2015oma} an optimal analysis method is discussed to search for a sharp emission line of SGWB, which 
  can increase the signal-to-noise ratio by up to a factor of seven.}
  Namely, our comparison here is only a rough one, but it is sufficient for our purposes. 
  \magenta{The same applies \green{to} the comparison with PTA. }
  }  
 }
}. 

\magenta{
Pulsar timing arrays (PTAs) have also been used to constrain GWs. Following \cite{Kuroyanagi:2014nba} 
we use the most stringent upper bound around $f=5.72\green{\times 10^{-9}}$Hz ($\sim \green{4\times 10^6}\rm{Mpc}^{-1}$), $\Omega_{\rm{GW}}\sim z_{\rm{eq}}^{-1}(4/11)^{1/3}{\cal P}_h\lesssim \umi{2}\times10^{-8}$, 
which leads to ${\cal P}_h\lesssim \umi{1}\times 10^{-4}$.
}

\red{Note that GW detectors \magenta{or PTA experiments} usually target GWs on a relatively limited frequency range, while cosmological methods like PBHs probe primordial GWs on a wide range of frequencies, 
and this is another advantage of PBHs in constraining primordial GWs \cyan{(see Fig.\ref{constraints})}. 
}

\textit{\magenta{Discussion}.-}
As already mentioned PBH constraints are applicable from $\sim \rm{Gpc}$ all the way down to $\sim 0.3$m 
if we assume the number of e-foldings during inflation is sixty. 
Note that the exclusion of an overproduction of smallest PBHs ($M_{\rm{PBH}}\lesssim 10^5g$) depends on the assumption that 
stable Planck mass relics are left over at the end of Hawking evaporation, which contribute to cold dark matter (see \cite{MacGibbon:1987my}, \cite{Carr:2009jm} and references therein). 
The range of comoving scales corresponding to $M_{\rm{PBH}}\lesssim 10^5g$ is roughly $\lesssim 50$m, 
and so PBH upper bounds in this range depend on this assumption. 
If Planck mass relics are not formed, 
to what extent an overproduction of PBHs with $M_{\rm{PBH}}\lesssim 10^5g$ is cosmologically problematic is uncertain. 
Such an overproduction of smallest PBHs may lead to an early \red{matter-dominated era}, 
during which  PBH binaries are formed and emit GWs, 
or larger PBHs may form \red{due to merger taking place after} the collapse of perturbations of PBHs' density, 
thereby leaving observable traces \cite{Dolgov:2011cq}. 
Therefore, in principle one may still exclude such an \umi{overproduction} of smallest PBHs even without the left over of Planck mass relics 
to fully validate our upper bounds on smallest scales, 
though we do not discuss it in detail here. 

We can also constrain a blue tensor power spectrum of the following form:
\be
{\cal P}_h(k)=r {\cal P}_\zeta(k_{\rm{ref}})
\left(
\frac{k}{k_{\rm{ref}}}
\right)^{n_T},\quad (k<k_{\rm{max}}),
\ee
where ${\cal P}_\zeta$ is the dimensionless power spectrum of the curvature perturbation on uniform-density hypersurfaces and 
$r$ is the so-called tensor-to-scalar ratio, here defined at $k_{\rm{ref}}\blue{=0.01\rm{Mpc}^{-1}}$. 
Let us assume $r=0.2$ and ${\cal P}(k_{\rm{ref}})=2.2\times 10^{-9}$ following \cite{Kuroyanagi:2014nba} and 
also $k_{\rm{max}}\sim 1\rm{Gpc}^{-1}e^{60}\sim 3\rm{m}^{-1}$. 
If we simply require ${\cal P}_h(k_{\rm{max}})\lesssim \blue{0.04}$, 
we obtain $n_T\lesssim 0.3$, which is tighter than other constraints such as BBN \red{constraints} shown in \cite{Kuroyanagi:2014nba}. 
Once more, this PBH bound does not depend on the details of a potential early \red{matter-dominated era} phase nor an entropy production, 
on which constraints other than PBHs are sensitive \cite{Kuroyanagi:2014nba}. 
If we use a more secure but weaker constraint on PBHs of around $10^5g$ by their entropy production \cite{Carr:2009jm}, 
we may require ${\cal P}_h(20\rm{km}^{-1})\lesssim \blue{0.4}$, leading to $n_T\lesssim 0.4$, 
still tighter than other constraints. 
To conclude, PBHs can provide \red{important} \cyan{constraints} also on a blue spectrum, though a more careful analysis would be merited. 

We have imposed $\sigma_{\delta_r}\lesssim 0.1\delta_{r,\mathrm{th}}$ to obtain PBH constraints, eq.(\ref{PBH}). 
The factor of $0.1$ here is to ensure the sufficient rareness of formation of PBHs. 
Note that to discuss the probability of PBH formation one needs to know the PDF of induced radiation density perturbation, 
which is in general highly non-Gaussian since it is generated by second-order tensor perturbations. 
In \cite{prep} we assume Gaussianity of tensor perturbations on top of the delta-function power spectrum, 
and numerically confirmed the factor of roughly $0.1$ is indeed necessary to avoid an overproduction of PBHs (compare eq.(\ref{PBH}) and the PBH constraints in Fig.\ref{constraints}). 
Strictly speaking, observational constraints on PBHs depend on their mass \cite{Carr:2009jm} and so how rare PBH formation has to be also 
depends on their mass, hence the scale dependence of PBH constraints in Fig.\ref{constraints} \cite{prep}. 

This argument implies PBH constraints on tensor perturbations depend on the statistics of tensor perturbations, determining the statistics of induced density perturbation, 
just as PBH constraints on \textit{scalar} perturbations depend on the statistics of scalar perturbations \cite{Byrnes:2012yx}. 
If high-$\sigma$ realizations of tensor perturbations are suppressed (enhanced) in comparison to a Gaussian case, PBH constraints on tensor perturbations are tighter (weaker). 

\textit{Acknowledgement.-}
We \cyann{are grateful to} Jun'ichi Yokoyama for reading the manuscript, useful comments and continuous encouragement.
We also thank \cyann{Kazunori Kohri,} Kazunari Eda, Yuki Watanabe, Yousuke Itoh, Daisuke Yamauchi, 
Tsutomu Kobayashi, Masahide Yamaguchi, Takahiro Tanaka \cyann{and anonymous referees} for helpful comments.
This work was supported by JSPS Grant-in-Aid for Young Scientists (B) No.15K17632 (T.S.), 
MEXT Grant-in-Aid for Scientific Research on Innovative Areas 
"New Developments in Astrophysics Through Multi-Messenger Observations 
of Gravitational Wave Sources" No.15H00777 (T.S.) 
and Grant-in-Aid for JSPS Fellow No.25.8199 (T.N.).

\bibliography{ref.bib}

\end{document}